\documentclass[twoside,reqno]{lt8-proc}
\usepackage{epsfig,cite}
\usepackage{amssymb}
\usepackage{amsfonts}
\usepackage{times}
\usepackage{graphics}
\begin{document}
\sloppy \raggedbottom
\setcounter{page}{1}
\newcommand{\beq}{\begin{equation}}
\newcommand{\eeq}{\end{equation}}
\newcommand{\beqa}{\begin{eqnarray}}
\newcommand{\eeqa}{\end{eqnarray}}
\newcommand{\nn}{\nonumber \\}
\newcommand {\np}[1]{{\mbox{\textrm{:}\,}{#1}{\,\textrm{:}}} }
\def \e {\mathrm{e}}
\def \edge {\mathrm{edge}}
\def \eps {\varepsilon}
\def \k {\kappa}
\def \l {\lambda}
\def \la {\langle}
\def \ra {\rangle}
\def \s {\sigma}
\def \t {\tau}
\def \B {{\mathcal B}}
\def \C {{\mathbb C}}
\def \Cl {{\mathcal C}}
\def \F {{\mathbb F}}
\def \M {{\mathcal M}}
\def \R {{\mathbb R}}
\def \Z {{\mathbb Z}}
\def \ch {\mathrm{ch}}
\def \el {\mathrm{el}}
\def \qh {\mathrm{qh}}
\def \qp {\mathrm{qp}}
\def \z {\zeta}
\def \L {\underline{\Lambda}}
\def \D {\Delta}
\def \I {{\mathbb I}}
\def \Pf {\mathrm{Pf}}
\def \P {{\mathcal P}}
\def \Im {\mathrm{Im} \, }
\def \Re {\mathrm{Re} \, }
\def \sh {\mathrm{sh}}
\def \tx {\tilde{x}}
\def \mod {\ \mathrm{mod} \ }
\def \H {{\mathcal H}}
\def \uu {{\widehat{u(1)}}}
\bibliographystyle{utphys}
\newpage
\setcounter{figure}{0}
\setcounter{equation}{0}
\setcounter{footnote}{0}
\setcounter{table}{0}
\setcounter{section}{0}



\title{Monodromy analysis of the computational power of the Ising topological quantum computer}

\runningheads{L. S. Georgiev}{Monodromy Analysis of the Ising TQC}

\begin{start}


\coauthor{Andre Ahlbrecht}{2,3}, 
\author{Lachezar S. Georgiev}{1,2} and

\coauthor{Reinhard~F.~Werner}{2,3}

\address{Institute for Nuclear Research and Nuclear Energy, \\
Bulgarian Academy of Sciences, \\
72 Tzarigradsko Chaussee, 
1784 Sofia, Bulgaria}{1}

\address{Insitut f\"ur Mathematische Physik, Technische Universit\"at Braunschweig,
Mendelssohnstr. 3, 38106 Braunschweig, Germany}{2}

\address{Insitut f\"ur Theoretische Physik, Leibniz Universit\"at Hannover,
Appelstr. 2, 30167 Hannover, Germany}{3}


\begin{Abstract}
We show that all quantum gates which could be implemented  by braiding of Ising anyons in the 
Ising topological quantum computer preserve the $n$-qubit Pauli group. Analyzing the structure 
of the Pauli group's centralizer, also known as the Clifford group, for $n\geq 3$ qubits, we prove 
that the image of the braid group is a non-trivial subgroup of the Clifford group and therefore 
not all Clifford gates could be implemented by braiding. We show explicitly the Clifford gates 
which cannot be realized by braiding estimating in this way the ultimate computational power 
of the Ising topological quantum computer.
\end{Abstract}
\end{start}

\section{Introduction}
Quantum computers are expected to be much more powerful than the classical supercomputers
due to a combination of quantum phenomena such as coherent superpositions, entanglement and 
paralelism \cite{nielsen-chuang}. 
Topological Quantum Computers are a class of quantum computers in which  information is encoded
in non-local topological quantum numbers, such as the anyon fusion channels, and quantum gates are 
implemented by anyon braiding protecting in this way quantum information processing from noise
\cite{kitaev-TQC,sarma-RMP}. 

In this paper we will analyze the properties of the monodromy subgroup of the braid group for the Ising 
anyon TQC and will demonstrate that all quantum gates which can be executed by braiding are Clifford 
gates \cite{nielsen-chuang,dam-howard}, that are important for fault-tolerant quantum computation,
while not all Clifford gates could be implemented by braiding Ising anyons.
\subsection{What are anyons?}
Anyons are particle-like collective excitations, which are  supposed to exist in strongly correlated two-dimensional 
electron systems, that may carry fractional electric charge (measured in shot-noise experiments with 
Laughlin quantum Hall states) and obey exotic exchange statistics: due to the properties of the two-dimensional 
rotation group $SO(2)$  the statistics under exchange of identical particles  is governed by representations 
of the braid group rather than the permutation group. The Abelian anyon many-body states belong to 
one-dimensional representations of the braid group and acquire nontrivial phases when neighboring anyons 
are exchanged. If the many-body states belong to a higher-dimensional representation of the braid group 
the corresponding particles are called non-Abelian anyons, or plektons, and   the results of particle exchanges
are not only phases but could also be more general unitary transformations  \cite{sarma-RMP,stern-review}.

Anyons might be observable in fractional quantum Hall samples as well as in
high-temperature superconductors  \cite{stern-review} and cold atoms in optical lattices (intersecting laser beams).
\subsection{Fusion paths: labeling anyonic states of matter}
The fact that the exchanges of non-Abelian anyons may generate non-trivial matrices acting on the many-body state
implies that in fact this state must belong to a degenerate multiplet of states with many anyons at fixed positions.
Therefore the anyon's positions and quantum numbers  \textit{are not sufficient for specifying a multi-anyon state},
i.e., some additional non-local information  is necessary. It appears that in order to specify the state
of many non-Abelian anyons we need to fix the fusion channels of any two neighbors. 
The reason is that the same multi-anyon configuration may correspond to different independent states (CFT blocks) 
because of the possibility of multiple fusion channels.
Consider, the process of fusing two anyons of type ``$a$" and ``$b$", represented by some operators $\Psi_a$ 
and $\Psi_b$ . The result is expressed by the following fusion rule
\[
\Psi_a \times   \Psi_b =   \sum\limits_{a=1}^{g}   N_{ab}^c   \Psi_c 
\]
where $N_{ab}^c$ are the fusion coefficients. There are two classes of anyons:
\begin{itemize}
\item {Abelian:} 
$\forall a,b \ \exists ! c$ such that  $N_{ab}^c \neq 0$
\item {non-Abelian:} 
if for some $a$ and $b$ $N_{ab}^c \neq 0$ for more than one $c$
\end{itemize}
In TQC we are interested in information encoding for non-Abelian anyons:
because by definition there are more than 1 fusion channels we could encode information in the 
index of the  fusion channel. 

For example, for Ising anyons $\Psi_I(z)=\s(z)$, represented by the chiral spin filed operator 
of CFT dimension $\D=1/16$, which are characterized by the fusion rule 
\[
\s \times \s = \I +\psi ,
\]
where $\psi$ is the Majorana fermion,
information is encoded as follows: if a pair of $\s$ fields is in the vacuum fusion channel we call 
the state $|0\ra$, while if it is  in the Majorana channel we call  the state $|1\ra$
\beqa
|0\ra=(\s,\s)_{\I}\quad\ &\longleftrightarrow& \quad \s \times \s  \to \I \nn
|1\ra=(\s,\s)_{\psi}\quad &\longleftrightarrow& \quad \s \times \s  \to \psi  \nonumber 
\eeqa
(the subscript of a pair denotes its fusion channel).
The important point here is that fusion channel is independent of the fusion process 
details--it is a topological quantity. It is also non-local because the fusion channel is independent 
of the anyon separation and is preserved even for large separation of anyons. 
Finally, it is robust and persistent--if we fuse two particles and then split them again the fusion channel 
does not change. Therefore, the message is that, in addition to the positions and all local quantum numbers 
of the anyons, a multi-anyon state  could be unambiguously specified 
by the \textit{fusion path}, i.e., the concatenation of elementary fusion channels for each neighbors
in an array of anyonic fields at fixed positions. This is conveniently  presented in the form of Bratteli 
diagrams (see, \cite{clifford} for more details).
\subsection{Quantum gates: adiabatic transport of anyons}
As we mentioned above, we intend to execute quantum gates over our quantum register by
adiabatic exchange of anyons. The adiabatic approximation requires that the system 
has a gapped Hamiltonian, i.e.,  a non-zero energy gap $\D$ to exist between the ground state 
and the  excitations.  The external parameters describing the transport are the anyons positions $R_1, \ldots, R_k$ 
in the plane (pinned by trapping potentials). 
We recall that we are interested in operations which keep the anyon's positions fixed
 so that final configuration is  at most a permutation of the anyons in the original one.
The elementary braiding processes are realized by taking one anyon adiabatically around its neighbor  
 on a large time scale $t\in [0,T]$ with $T\gg \D^{-1}$ 
(for example, if the energy gap of the $\nu=5/2$ FQH state is $\D=500$mK then the minimum time interval 
is $T_{\min}\sim 10^{-10}$s). The adiabatic theorem for a non-degenerate ground state implies that
  if  the system is initially in the ground state then the final state after executing a complete loop in the 
parameter space, is  up to phase again the ground state
\beq\label{adiabatic}
\psi^{(R_1, \ldots, R_k)}_f(z_1,\ldots,z_N)=
\e^{i\phi} \psi^{(R_1, \ldots, R_k)}_i(z_1,\ldots,z_N),
\eeq
where $z_i$ are the coordinates of the electrons while $R_j$ are the coordinates of the anyons in the plane.
However, if the ground state is degenerate and separated by a gap from the excited states, then the adiabatic 
theorem implies that the final state after traversing a complete loop is again a ground state from the same 
degenerate multiplet but may be different from the original one. In this case the phase $\e^{i\phi}$ in 
(\ref{adiabatic}) must be replaced by a unitary operator acting on the multiplet \cite{sarma-RMP}.

Ignoring the dynamical phase $\e^{i\frac{1}{\hbar}\int dt E({\bf R}(t))}$ contribution to $\e^{i\phi}$ we will focus 
on the (non-) Abelian Berry phase $\e^{i\alpha}$ defined by
\beq\label{alpha}
\alpha=\oint d{ \vec{\bf R}} \, \cdot  \, \la \psi( \vec{\bf R} ) | \vec{\nabla}_{\bf R} |\psi( \vec{\bf R} )\ra  ,
\quad \vec{\bf R} = (R_1, \ldots, R_k) 
\eeq
There are three contributions to the Berry phase (\ref{alpha}), which have different status with respect to the 
topological protection:
\begin{itemize}
\item{geometrical phase which is of the type of  the Aharonov--Bohm phase }
\item{topological phase (quasiparticle statistics, independent of the geometry)}
\item{monodromy of CFT wave functions with anyons}
\end{itemize}
The geometrical phase is proportional to the area of the loop and is not topologically protected. 
When the many-body states are described by chiral
CFT wave functions which are also orthonormal, the Berry connection is trivial \cite{read-viscosity}.
Because of this, the entire effect of the adiabatic 
transport is given by the explicit monodromy of the multi-valued CFT correlators. Therefore, when we construct 
quantum gates for TQC, we can directly deal with the braid generators and forget about the Berry connection 
induced by the adiabatic transport.
\section{$n$ Ising qubits: $2n+2$ Ising anyons on antidots}
\begin{figure}[htb]
\centering
{\includegraphics*[bb=35 595 390 685,width=8.5cm]{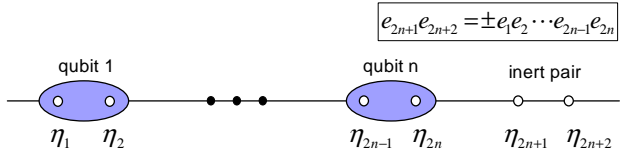}}
\caption{The $n$-qubit encoding scheme: $e_i =\pm 1$ is the chiral fermion parity of 
the $i$-th anyon field $\s_{e_i} (\eta_i)$  so that $c_i=e_{2i-1} e_{2i}$, for $1\leq j\leq n+1$. The
sign $\pm$ in the box equation corresponds to the parity of the spinor 
representation of the braid group \label{fig:n-qubits}}
\end{figure}

Because the dimension of the computational space, spanned by the Pfaffian wave functions with $2n+2$ Ising 
anyons at fixed positions $\eta_1, \ldots \eta_{2n+2}$ in the plane \cite{nayak-wilczek}, is $2^n$ we could use 
the many-body states represented by these wave functions to realize $n$ Ising qubits.
Our qubit encoding scheme is roughly that we use one pair of $\s$ fields to represent one qubit, whose state
is $|0\ra$ if the pair is in the fusion channel of the vacuum or $|1\ra$ if it is in the Majorana channel.
However, since we would like to represent qubits by chiral CFT correlation functions of Ising $\s$ fields, 
which are non-zero only if the total fermion parity inside the correlator is trivial, we need one extra pair of 
$\s$ fields, which is inert from the viewpoint of TQC but compensates if necessary total parity of the encoded 
qubits, i.e.,  our one-qubit states could be written as 4-pt correlators
\[
|c_1 \ra =  \la (\s \s)_{c_1} (\s \s)_{c_0} \ra_{\mathrm{CFT}}, \quad  \mathrm{with} \quad  c_0=c_1,
\]
where the subscript of the pair  $(\s \s)_{c}$ denotes its fusion channel $c$.
Similarly, we can represent $n$ qubits as a correlator of  $(n+1)$ pairs of Ising anyons $\s$ 
 where the last pair compensates the total fermion parity
\[
|c_1, \ldots, c_i, \ldots, c_n \ra \to \la (\s \s)_{c_1}
\cdots (\s \s)_{c_i} \cdots (\s \s)_{c_n}   (\s \s)_{c_0}\ra_{\mathrm{CFT}},
\]
with $c_i=\pm$ being the fermion parity of the $i$-th pair of $\s$ fields.
This encoding scheme is illustrated in Fig.~\ref{fig:n-qubits}

\subsection{Braid matrices: multi-anyon wave-function approach}
The explicit form of the generators of the Ising representation of the braid group $\B_{2n+2}$
has been conjectured in \cite{nayak-wilczek} to coincide with the finite subgroup of the 
$\pi/2$-rotations from SO$(2n+2)$ (more precisely, with one of the spinor representations of its double cover 
Spin$(2n+2)$). More robust results have been obtained in \cite{ivanov,read-JMP,franko}.
However, a more natural approach, which is based on the direct computation of the braid matrices 
by analytic continuation of the multi-anyon Pfaffian wave functions, using the different operator product expansions
of the Ising $\s$ fields in the Neveu--Schwarz and Ramond sectors of the Ising model, has been exploited 
in \cite{ultimate}, which allowed to write explicitly the  generators  $B_1^{(2n+2,\pm)}, \ldots B_{2n+1}^{(2n+2,\pm)}$ 
of the positive/negative parity representations (denoted by $\pm$ in the superscript) of $\B_{2n+2}$ 
for arbitrary number $n$ of qubits \cite{ultimate}
\beq\label{B_2j-1}
B_{2j-1}^{(2n+2,\pm)}= \underbrace{\I_2\otimes \cdots \otimes \I_2}_{j-1} \otimes 
\left[ \matrix{1 & 0 \cr 0 & i}\right] \otimes \underbrace{\I_2\otimes \cdots \otimes \I_2}_{n-j} ,\ \ 
\textrm{for} \ \  1\leq j\leq n,
\eeq
\beq\label{B_2j}
B_{2j}^{(2n+2,\pm)}= \underbrace{\I_2\otimes \cdots \otimes \I_2}_{j-1} \otimes 
\frac{\e^{i\frac{\pi}{4}}}{\sqrt{2}}
\left[ \matrix{1 & 0 & 0 & -i \cr 0 & 1 & -i & 0 \cr 0 & -i & 1 & 0 \cr -i & 0 & 0 & 1 }\right]  
\otimes  \underbrace{\I_2\otimes \cdots \otimes \I_2}_{n-j-1} ,
\eeq
for $n\geq 2$ and  $1\leq j\leq n-1$, as well as
\beq\label{B_2n}
B_{2n}^{(2n+2,\pm)}= \underbrace{\I_2\otimes \cdots \otimes \I_2}_{n-1} \otimes 
\frac{\e^{i\frac{\pi}{4}}}{\sqrt{2}}
\left[ \matrix{1 & -i \cr -i & 1  }\right] .
\eeq
The last braid generators $B_{2n+1}^{(2n+2,\pm)}$ of the $(\pm)$-parity representations  of  $\B_{2n+2}$
cannot be written in a similar form for general $n$ because they do not have a tensor product structure.
Yet, these diagonal matrices can be determined using Eq. (32) in \cite{clifford},  the results after
Eq.~(24) in \cite{equiv} and Proposition~2 in \cite{ultimate}
\beq\label{B_2n+1}
B_{2n+1}^{(2n+2,\pm)}=\frac{\e^{i\frac{\pi}{4}}}{\sqrt{2}} \left( \I_{2^n}\mp  
i \underbrace{\s_3\otimes \cdots \otimes \s_3}_{n} \right) ,
\eeq
where $\s_3$ is the third Pauli matrix.
The above equations (\ref{B_2j-1}),  (\ref{B_2j}), (\ref{B_2n}) and (\ref{B_2n+1}) 
provide the most explicit and compact form of the generators
of the two representations of $\B_{2n+2}$ with opposite fermion parity.  Because
the Berry connection for adiabatic transport of Ising anyons is trivial \cite{read-viscosity},
 these braid matrices can be ultimately used  to implement topologically protected quantum gates
 by  adiabatic transport in the Ising TQC.
\section{Pauli group for $n$ qubits: quantum correctable errors}
One fundamental structure in any quantum information processing platform is the group of Pauli matrices, or Pauli 
gates, which are part of the definition of the computational basis. They are important not only because they represent
a group of essential quantum operations but also because of the quantum error correction specifics.
The point is that there are two independent types of errors which can be considered as deviations of the 
point, representing the qubit,  of the Bloch sphere. Bit-flip ($\s_x$) errors are deviations along the meridians 
while phase-flip ($\s_z$) errors are deviations along the parallels of the sphere. 
While it is obvious that arbitrary errors can be decomposed into bit-flip and phase-flip errors, unlike in classical 
error correction, the errors which could compromise a qubit are continuous quantities. Fortunately, it appears that
this continuum of (arbitrary) errors can be corrected by correcting only a discrete subset of those errors: e.g., as in
the Shor code \cite{nielsen-chuang}.
Due to this virtue of the quantum correctability it is sufficient to consider and correct errors which belong to the Pauli
group. The $n$-qubit Pauli group is defined as the finite group containing all Pauli matrices $\s_j$ acting on any 
of the qubits, including phases of $\pm i$
\beq\label{Pauli}
\P_n=\left\{ i^{m}\sigma_{\alpha(1)} \otimes \cdots \otimes \sigma_{\alpha(n)}\, \left| \quad \alpha(j), \ m\in \{0,1,2,3\} \right.  \right\},
\eeq
(with $\s_0=\I_2$).
The projective Pauli group  is isomorphic to $\Z_2^{2n}$ (see Eq.~(17) in  \cite{clifford}), and its 
center is $\Z_4$, so that the order of the $n$-qubit Pauli group 
(\ref{Pauli}) is $|\P_n|=2^{2n+2}$.
\subsection{$n$-qubit Clifford group: symplectic description }
Because of the fundamental importance of the Pauli group its stabilizer also plays a very important role.
By definition the  stabilizer of the $n$-qubit Pauli group, known as the $n$-qubit Clifford group, is the group of all 
unitary $2^n \times 2^n$ matrices which preserve the Pauli group
\beq\label{Clifford}
\Cl_n=\left\{ U \in SU(2^n)\ | \ U^*\cdot \P_n \cdot U \subset \P_n \ \right\}. 
\eeq
The fact that the Clifford unitaries commute with the Pauli operators makes them ideal for quantum error 
correction because they do not introduce new errors while correcting the existing ones.

The Pauli group is naturally a subgroup of the Clifford group (\ref{Clifford}), i.e.,  $\P_n \subset \Cl_n$. 
The Clifford group is infinite, however, the projective Clifford group $[\Cl_n] \equiv \Cl_n/Z$, where $Z$ is 
its center, is finite.
Furthermore, there is an interesting isomorphism \cite{clifford} between the projective Clifford group $[\Cl_n]$ factorized 
by the projective Pauli group $[\P_n] \equiv \P_n/\Z_4$ and the symplectic group $Sp_{2n}(2)$  
(the group of symplectic $2n\times 2n$ matrices with elements $0$ and $1$)
\[
	[\Cl_{n}]/[\P_n] \simeq Sp_{2n}(2).
\]
This isomorphism allows us to compute the order of the projective Clifford group 
using the known order of the symplectic group $Sp_{2n}(2)$ \cite{wilson} (see also Appendix~A in 
\cite{clifford}  as well as the order of the projective Pauli group
\beq\label{order-Cl}
	\left| \Cl_{n}/Z \right|=   2^{n^2+2n} \prod_{j=1}^n (4^j-1) \, .
\eeq
This result will be important when we try to estimate the computational power of the Ising TQC.
\subsection{Braiding gates as Clifford gates}
A very useful observation in the context of TQC with Ising anyons is that the $n$-qubit Pauli group completely 
coincides with the monodromy subgroup of the braid group representation $\B_{2n+2}$
\beq \label{isomorphism}
	\P_n \equiv \mathrm{Image}\left(\M_{2n+2} \right).
\eeq
Then, because the monodromy group is a normal subgroup of the braid group,  $\M_{2n+2} \subset \B_{2n+2}$, i.e.,
\[
\forall b\in \B_{2n+2} , \forall m \in \M_{2n+2} :  \quad b^{-1} \, m \, b \in \M_{2n+2},
\]
it follows that all braiding gates are Clifford gates, i.e., the image of the braid group $\B_{2n+2}$ is a subgroup 
of the Clifford group for $n$ Ising qubits
\[
	 \mathrm{Image}(\B_{2n+2}) \subset \Cl_n    .
 \]
 To prove this, notice first that
$\P_n\subset \mathrm{Image}\left(\M_{2n+2} \right)$ because all Pauli gates could be expressed in terms of
the squares of the elementary braid generators, which belong to the monodromy group, i.e., 
for $1\leq j\leq 2n+1$, we have for the spinor representations generators $R_{j}^{(n+1,\pm)}$ of $\B_{2n+2}$ 
(which have been proven  in Proposition~2 in \cite{ultimate} to be equivalent to our representations with 
generators $B_{j}^{(2n+2,\pm)}$ derived directly from the wave-function)
\beq\label{R_2j-1}
\left(R_{2i-1}^{(n+1,+)} \right)^2 =
\underbrace{\I_2\otimes  \cdots \otimes \I_2}_{i-1} \otimes \sigma_3 \otimes
\underbrace{\I_2\otimes \cdots \otimes \I_2}_{n-i},
\eeq
\beq\label{R_2j}
\left(R_{2i}^{(n+1,+)} \right)^2 =
\underbrace{\I_2\otimes \cdots \otimes \I_2}_{i-1} \otimes \sigma_2\otimes \sigma_2 \otimes
\underbrace{\I_2\otimes  \cdots \otimes \I_2}_{n-i-1},
\eeq
\beq\label{R_2n}
\left(R_{2n}^{(n+1,\pm)} \right)^2 =
\mp\underbrace{\sigma_3\otimes \cdots \otimes \sigma_3}_{n-1} \otimes \sigma_1
\eeq
\beq\label{R_2n+1}
\left(R_{2n+1}^{(n+1,\pm)} \right)^2 = \pm
\underbrace{\sigma_3\otimes \cdots \otimes \sigma_3}_{n}          .
\eeq
Therefore the Pauli gates could be explicitly written in terms of monodromies
\[
\sigma_2^{(n)} = i \left(R_{2n}^{(n+1,+)}\right)^2 \left(R_{2n+1}^{(n+1,+)}\right)^2\, ,
\]
\[
\sigma_2^{(n-j)} = i \left(R_{2n-2j}^{(n+1,+)}\right)^2 \s_2^{(n-j+1)} , \quad
1\leq j\leq n-1.
\]
On the other hand, as shown in \cite{clifford}, the monodromy generators $A_{ij}$, with 
$1\leq i<j \leq 2n+2$, which can be presented in the form 
\[
A_{ij}\equiv U_{ij}^{-1} B_{i}^2 U_{ij} , \quad \mathrm{where} \quad U_{ij}= \prod_{k=i+1}^{j-1} B_k,
\]
can be expressed in terms of the Pauli generators  \cite{clifford} due to Eqs. (\ref{R_2j-1}), 
(\ref{R_2j}), (\ref{R_2n}) and (\ref{R_2n+1}) because
\[
A_{kl}^{\pm}= - (-i)^{l-k+1} \left(R_k^{\pm}\right)^2\left(R_{k+1}^{\pm}\right)^2 \cdots 
\left(R_{l-2}^{\pm}\right)^2\left(R_{l-1}^{\pm}\right)^2,
\]
where $1\leq k < l \leq 2n+2$ and we omitted  the $(n+1)$ in the superscripts. 
This completes the proof of the statement that all quantum gates which can be implemented by braiding of 
Ising anyons are actually Clifford gates.
\subsection{Orders of the image of the braid group and the Clifford group}
We saw in the previous subsection that all braiding gates are Clifford gates.
Unfortunately, the converse is not true--not all Clifford gates could be implemented by
braiding of Ising anyons. To see this let us compare the order of the Clifford group (\ref{order-Cl}) with that 
of the braid group $\B_{2n+2}$ in the Ising representation, which is given by \cite{read-JMP,clifford}
\beq \label{order-B}
\left| \mathrm{Image}\left( \B_{2n+2}\right)\right| = 2^{2n+2}  (2n+2)!, \quad n\geq 2,
\eeq
and the order for $n=1$ (including the center) is  $\left| \mathrm{Image}\left( \B_{4}\right)\right| = 96$, 
see Ref.~\cite{TQC-NPB}.
Using Eqs.~(\ref{order-B}) and (\ref{order-Cl}) we compare in Table~\ref{tab:1} the orders of the projective 
braid and Clifford groups  for a few qubits.
\begin{table}[htb]
\caption{Comparison of the orders of the projective image of the braid group $\B_{2n+2}$ and the projective 
Clifford group $\Cl$ for $n$ Ising qubits \label{tab:1}}
 \begin{tabular}{l||r|r|r|r|r}
 	\hline
	$n$ & 1 & 2 & 3 & 4 & 5  \\
\hline \hline
$\left|\B_{2n+2}/ \Z_4\right|$& 24 & 11520 & 2580480 & $\propto 0.9\times 10^{9}$
& $\propto 0.5\times 10^{12}$ \\
\hline
$|\P\Cl_n|$ & 24 &  11520 & 92897280 & $\propto 1.2\times 10^{13}$ &
$\propto 2.5\times 10^{19}$\\
\hline
	\end{tabular}
\end{table}
It is obvious that the order of the Clifford group grows faster with the number of the qubits than the image
of the braid group.  
\begin{figure}[htb]
\centering
\includegraphics*[bb=55 30 580 440,width=9cm]{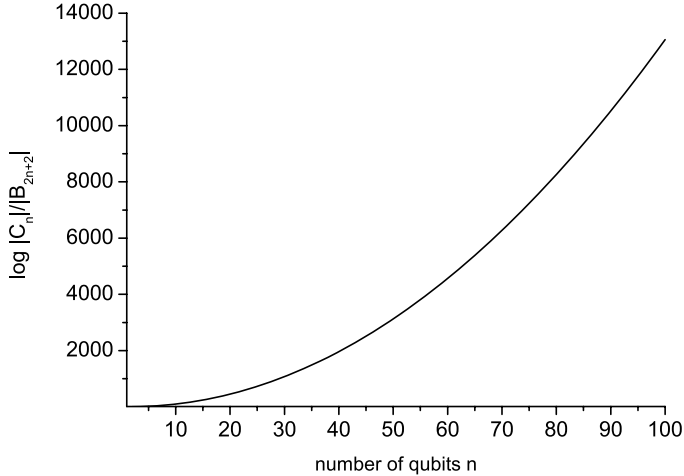}
\caption{Logarithm of the ratio of the order of the Clifford group and that of the image of the braid group
as a function of the number of qubits $n$.  \label{fig:log}}
\end{figure}
In Fig.~\ref{fig:log} plot the logarithm of the ratio of the order of the projective Clifford group and 
the order of the image of the braid group, corresponding to $n$ qubits, as a function of the number of qubits $n$.
This logarithm still grows quadratically with $n$ which means that the order of the Clifford group
grows exponentially faster with $n$ than the order of the image of the braid group.
As can be seen from Table~\ref{tab:1}, the only exceptions are $n=1$ and $2$ for which the entire 
Clifford group could be implemented by braiding \cite{TQC-NPB}.
Therefore, it is not possible to realize all Clifford gates for $n\geq 3$ by braiding of Ising anyons.
\section{Conclusions}
In this paper we have demonstrated that the $n$-qubit Pauli group for TQC with Ising anyons coincides exactly
with the representation of the monodromy subgroup  $\M_{2n+2}$ of the braid group $\B_{2n+2}$ describing 
the exchanges of $2n+2$ Ising anyons. This implies that all braiding gates are actually Clifford gates, which is important for fault-tolerant quantum computation. However, not all Clifford gates are realizable by braiding only.
The typically   missing Clifford gates from the braid realization are the SWAP gates \cite{clifford}
which simply exchange the quantum states of two qubits in an $n$-qubit quantum register \cite{nielsen-chuang}.
This is another limitation of the Ising-anyon topological quantum computer, which has already been known
\cite{freedman-larsen-wang-braid} to be
non-universal for quantum computation since some non-Clifford gates are not realizable by braiding.
\section*{Acknowledgments}
We would like to thank Lyudmil Hadjiivanov, Holger Vogts,  Sergey Bravyi, Volkher Scholz and 
Johannes Guetschow  for useful discussions.
L.S.G. has been supported as a Research Fellow by the Alexander von Humboldt
foundation. This work has been partially supported by the BG-NCSR under Contract
No. DO 02-257.
\bibliography{Z_k,my,TQC,FQHE}

\end{document}